\documentstyle[12pt]{article}
\newcommand{\PRL}{{\em Phys. Rev. Lett.} }
\newcommand{\PR}{{\em Phys. Rev.} }
\newcommand{\PL}{{\em Phys. Lett.} }
\newcommand{\JMP}{{\em J. Math. Phys.} }
\newcommand{\NP}{{\em Nucl. Phys.} }
\newcommand{\de}{\delta}
\newcommand{\ph}{\varphi}

\newcommand{\om}{\omega}
\renewcommand{\inf}{\mathop{\rm inf}}
\newcommand{\ldel}{\langle}
\newcommand{\rdel}{\rangle}
\newcommand{\bra}[1]{\ldel\,#1\,|}
\newcommand{\ket}[1]{|\,#1\,\rdel}
\newcommand{\braket}[2]{\ldel\,#1\,|\,#2\,\rdel}

\newcommand{\be}{\begin{equation}}
\newcommand{\ee}{\end{equation}}
\newcommand{\bea}{\begin{eqnarray}}
\newcommand{\eea}{\end{eqnarray}}
\newcommand{\lb}{\label}

\begin{document}

\begin{titlepage}
\begin{flushright}
Z\"urich University Preprint\\
ZU-TH 7/94
\end{flushright}
\vspace{3 ex}
\begin{center}
{\bf
{\Large\bf Instability of Gravitating Sphalerons}
}
\vskip .8in
{\bf P. Boschung, O. Brodbeck, F. Moser, N. Straumann and M. Volkov}\\
\vskip 0.8cm
{\it Institute for Theoretical Physics, University of Z\"urich, }\\
{\it Winter\-thur\-er\-stras\-se\nobreak\ 190, 8057 Z\"urich, Switzerland}
\end{center}
\vskip 3cm
\begin{abstract}
We prove the instability of the gravitating regular sphaleron solutions of the
$SU(2)$ Einstein-Yang-Mills-Higgs system with a Higgs doublet, by studying the
frequency spectrum of a class of radial perturbations. With the help of a
variational principle we show that there exist always unstable modes. Our
method has the advantage that no detailed knowledge of the equilibrium solution
is required. It does, however, not directly apply to black holes.
\end{abstract}
\end{titlepage}
\section{Introduction}
\label{Intro}

When gravity is coupled to nonlinear field theories, such as Yang-Mills fields
or nonlinear $\sigma$-models, interesting and surprising new types of
particle-like and black hole solutions have turned out to exist.
Among the regular solutions the most interesting ones are those for which
gravity is essential.
The first example of this type was found by Bartnik and McKinnon \cite{BART}
for the Einstein-Yang-Mills (EYM) system.
For the same model several authors \cite{VOLK} discovered later the colored
black hole solutions which showed that the classical uniqueness theorem for the
Abelian case does not generalize.
The existence of both types of solutions which meanwhile has been established
rigorously \cite{SMO1,SMO2,SMO3}, and also because they were proven to be
unstable \cite{STR1,STR2,ZHO1,ZHO2}, led to a search for corresponding
solutions of other related field theories.
It turned out, for instance, that the Einstein-Skyrme (ES) system has black
hole solutions with hair which are at least linearly stable
\cite{DROZ,HEU1,HEU2,HEU3}.
(For a numerical investigation of nonlinear stability, see ref. \cite{HEU3}.)
Several authors looked at other models, notably the $SU(2)$
Einstein-Yang-Mills-Higgs (EYMH) theory with a Higgs triplet
\cite{LEE,BREI,AICH}, as well as the EYM-dilaton theory \cite{LAVR}, and found
in some cases other linearly stable black hole solutions.
Interesting black hole solutions have recently been found numerically for the
$SU(2)$ EYMH-theory with a Higgs doublet \cite{GREE}, as in the standard
electroweak model.
These ''sphaleron black holes'' were suspected to be unstable, but this
question has so far not yet been analyzed.

In the present paper we show that the regular sphaleron solutions are unstable,
but the stability issue for the corresponding black holes remains unsettled.
(Some partial results will be mentioned in the concluding section.)
Our proof proceeds along the following lines. First, we show that the frequency
spectrum of a class of radial perturbations is determined by a coupled system
of  radial "Schr\"odinger equations", which will be derived in section
\ref{PULS}  by linearizing the basic equations (given in section \ref{Basic})
around an equilibrium solution. Bound states of this Schr\"odinger problem
correspond to exponentially growing modes. Using the variational principle for
the ground state it is then proven in section \ref{Inst} that there exist
always unstable modes if the soliton is a purely magnetic solution of the EYMH
equations. We show this with the help of a judiciously choosen one--parameter
familiy of trial perturbations. Unfortunatly, this cannot be applied directly
to the black hole solutions, because of problems related to the boundary
conditions at the horizon.
We have the suspicion that some of the black hole configurations might be
(linearly) stable. The reasons for this will be mentioned in the final section
\ref{Remarks}.

The instability proof presented below is quite powerful for solitons, because
no detailed knowledge of the equilibrium solution is required. It has recently
been generalized by some of us \cite{BROD2} to the EYM system for arbitrary
gauge groups.

\section{Basic Formulae}
\label{Basic}
Since we are interested in the stability of spherically  symmetric black hole
and soliton solutions of the EYMH theory, we restrict ourselves to spherically
symmetric fields.

The metric is parametrized in the usual manner
\be
\lb{NS1}
g = - NS^2 dt^2 + N^{-1}dr^2 + r^2(d\vartheta^2 + \sin^2\vartheta d\varphi^2).
\ee
Instead of $N$ which is only a function of $t$ and $r$, we use also the metric
function $m(t, r)$ (mass function), defined by $N= 1 - 2m/r$ (we set $G=1$).

For the gauge potential $A$ there are many gauge equivalent ways to parametrize
the general spherically symmetric gauge
potential. (A systematic analysis for an arbitrary gauge group can be found in
\cite{BROD}.) A convenient representation is
\be
\lb{NS3}
A = a_0 \tau_r dt + a_1 \tau_r dr + (\om - 1) \, \left[ \tau_{\varphi}
d\vartheta - \tau_{\vartheta} \sin\vartheta d\varphi \right] + \tilde\omega \,
\left[ \tau_{\vartheta} d\vartheta - \tau_{\varphi} \sin\vartheta d\varphi
\right],
\ee
where $a_0,\, a_1,\, \omega$ and $\tilde\omega$
are functions of $t$ and $r$, and
$\tau_r$,
$\tau_{\vartheta}$,
$\tau_{\varphi}$
 are the spherical $SU(2)$ generators, defined by
$\tau_r = \vec\tau \cdot \vec e_r$,
$\tau_{\vartheta} = \vec\tau \cdot \vec e_{\vartheta}$,
$\tau_{\varphi}=\vec\tau\cdot \vec e_{\varphi}$,
 with the normalization $\vec\tau = \vec\sigma / 2i$
($\vec\sigma$: Pauli matrices) and
$\vec e_r, \vec e_{\vartheta}, \vec e_{\varphi}$ the unit vectors in the
directions of the coordinates $r$, $\vartheta$, $\varphi$.

The Higgs doublet can always be represented as a $2 \times 2$ matrix of the
form $\Phi = (\phi + i \vec\sigma \cdot \vec\psi)$,
$\phi$ and $\vec\psi$ real, which transforms under the gauge group by
left-multiplication.
A spherically symmetric Higgs field has the form
\be
\lb{NS4}
\Phi = \frac{1}{\sqrt{2}}\, (\phi \cdot {\bf 1} + i \psi \sigma_r).
\ee
It is now straightforward to compute the EYMH action for these fields.
 For the matter part, with the Lagrangian ${\cal L}_{\rm mat}$, we define the
effective radial-temporal Lagrangian, $L_{\rm mat}$, by the equation
\be
\lb{NS5}
\int {\cal L}_{\rm mat} \sqrt{-g}\, d^4x = \int L_{\rm mat} S dt\, dr.
\ee
One finds quite easily
\be
\lb{NS6}
L_{\rm mat} = - \frac{r^2}{4} f_{\mu\nu}f^{\mu\nu} + \frac{1}{NS^2} T  - NU -
P,
\ee
where $f_{\mu\nu}$                      $(\mu, \nu = 0,1)$ is the
two--dimensional field strength
\be
f_{\mu\nu} = \partial_{\mu} a_{\nu} - \partial_{\nu} a_{\mu},
\ee
and $U$, $T$ and $P$ are given by the following expressions in terms of the
complex matter variables $f = \omega + i\tilde\omega$, $h = \phi + i \psi$:
\be
T = \overline{D_0 f}D_0 f + \frac{r^2}{2} \overline{D_0 h} D_0 h,
\ee
\be
\lb{NS7}
U = \overline{D_1 f} D_1 f + \frac{r^2}{2} \overline{D_1 h} D_1 h,
\ee
\be
\lb{NS8}
P = \frac{(1-\bar f f)^2}{2r^2} + \frac{1}{4} \bar h h (1 + \bar f
f)-\frac{1}{2} {\rm Re} (\bar f h^2) + r^2 V(\bar h h).
\ee
Here, $V = \frac{\lambda}{4} (\bar h h - v^2)^2$ is the Higgs potential, and
the covariant derivatives are defined as $D_{\mu} f = (\partial_\mu - i a_\mu )
f$ and $D_{\mu} h = (\partial_\mu - i a_\mu / 2) h$ ($\mu, \nu = 0, 1$).
Note that $P$ depends only on the matter fields and $U$ and $T$ only on their
covariant derivatives.
(We have chosen the relative weight of the Yang-Mills and Higgs parts such that
annoying factors $4\pi$ do not enter in $L_{\rm mat}$. In this respect we
follow the conventions of ref.\ \cite{GREE}.)

The independent Einstein equations are
\bea
\lb{EM1}
m' &=& - \frac{r^2}{4} f_{\mu\nu}f^{\mu\nu} + \frac{1}{NS^2} T + NU + P,\\
\lb{EM2}
\dot m &=& 2N \mbox{Re} \left\{ \overline{D_1 f} D_0 f + \frac{r^2}{2}
\overline{D_1 h} D_0 h \right\},\\
\lb{EM3}
(\ln S)' &=& \frac{2}{r} \left\{ \frac{1}{(NS)^2} T + U\right\},
\eea
and the Yang--Mills--Higgs equations reduce to
\bea
\lb{YM1}
\partial_\mu (r^2 S f^{\mu\nu} ) &=& 2 S \, \mbox{Im} \left\{ f \overline{D^\nu
f} + \frac{r^2}{4} h \overline{D^\nu h} \right\},\\
\lb{YM2}
\frac{1}{S} D_\mu (S D^\mu f) &=& \frac{\bar ff-1}{r^2}f + \frac{\bar hh}{4} f
- \frac{1}{4} h^2,\\
\lb{HI}
\frac{1}{S} D_\mu (S \frac{r^2}{2} D^\mu h) &=& \frac{\bar ff + 1}{4} h  -
\frac{1}{2} f \bar h + \frac{r^2}{2} \lambda ( \bar hh - v^2)h.
\eea

\section{The Pulsation Equations}
\lb{PULS}
We now proceed to a linear stability analysis of a given equilibrium solution,
which is assumed to be a static, regular, asymptotically flat and purely
magnetic solution of the coupled EYMH equations. We choose the temporal gauge
by setting $a_0 = 0$. This can always be obtained with a gauge transformation
of the form $\exp(\tau_r \alpha)$. In this gauge the linearized equations will
lead to the standard Schr\"odinger eigenvalue problem for the pulsation
frequencies.
({}From now on, the symbols $f$, $h$ etc.\ refer to the equilibrium solution,
and
$\de f$, $\de h$ etc.\ denote their time dependent perturbations.)
As a first step we consider the Einstein equations, and obtain the following
linearized equations:
\bea
\lb{PB1}
\de m' &=& N \de U - \de m \frac{2}{r} U + \de P,\\
\lb{PB2}
\de \dot m  &=& 2N \mbox{Re} \left\{ \de \dot f \bar f' + \frac{r^2}{2} \de\dot
h \bar h' \right\},\\
\lb{PB3}
\de (\ln S)' &=& \frac{2}{r} \de U ,
\eea
where
\bea
\lb{PB4}
\de U &=& 2 \mbox{Re} \left\{ \bar f' \de f' + \frac{r^2}{2} \bar h' \de h'
\right\} + 2 \de a_1 \mbox{Im} \left\{ f \bar f' + \frac{r^2}{4} h \bar h'
\right\}, \\
\lb{PB5}
\de P &=& \left\{ \frac{\bar ff + 1}{2} + r^2 \lambda ( \bar hh - v^2) \right\}
\,\mbox{Re} ( \bar h \de h) \nonumber \\ &&+  \left\{ 2 \frac{\bar ff-1}{r^2} +
\frac{\bar hh}{2} \right\} \,\mbox{Re} (\bar f \de f) - \, \mbox{Re} ( \bar f h
\de h) - \frac{1}{2} \, \mbox{Re} (h^2 \de \bar f).
\eea

The perturbations of the matter fields $f$ and $h$ can be decomposed into
"real" and "imaginary" parts as follows.

Assume that $f = \omega $ and $h = \varphi$ with $\omega$ and $\varphi$ real,
and let us decompose the perturbations as
\bea
\lb{PB6}
\de f &=& \de \rho + i \de \chi,
\\
\lb{PB7}
\de h &=& \de \sigma + i \de \xi ,
\eea
with $\de \rho$, $\de \sigma$, $\de \chi$ and $\de \xi$ real.
We will call $\de \rho$ and $\de \sigma$ the "real" parts of the perturbations
and $\de \chi$ and $\de \xi$ their "imaginary" parts.

{}From the full  linearized system, one can easily conclude that the real and
the imaginary parts decouple. We are interested only in imaginary
perturbations, because we shall find instabilities within this class. In this
case the metric perturbations, $\de m$ and $\de S$,  vanish identically.
Indeed, from equation (\ref{PB2}) we conclude that $\de m$ is a function of $r$
alone. Using $\de U = 0$, $\de P = 0$ and an Einstein equation for the
equilibrium solution, equation (\ref{PB1}) leads then to the differential
equation $(\de m S)' = 0$. Together with the boundary conditions, $\de m(0) =
\de m(\infty ) = 0 $, this implies $\de m = 0$. $\de S = 0$ follows directly
from (\ref{PB3}) and the boundary conditions $\de S(0) = \de S(\infty ) = 0$.

Next we consider the linearized matter equations. For imaginary perturbations,
these can now be written in operator form as follows.
Let
\be
\lb{PB8}
\Psi = \left(
\begin{array}{l}
\de a_1 \\
\de \chi \\
\de \xi \\
\end{array}
\right), \qquad
H = \left(
\begin{array}{lll}
H_{aa} & H_{a\chi} & H_{a\xi} \\
H_{\chi a} & H_{\chi \chi} & H_{\chi\xi} \\
H_{\xi a} & H_{\xi \chi} & H_{\xi \xi} \\
\end{array}
\right),
\ee
with
\bea
\lb{PB10p}
H_{aa} &=& 2 (NS)^2 ( \om^2 + \frac{r^2}{8} \ph^2), \\
\lb{PB14p}
H_{\chi\chi} &=& 2 p_\ast^2 + 2NS^2 \left\{ \frac{\om^2-1}{r^2} +
\frac{\ph^2}{4} \right\}, \\
\lb{PB17p}
H_{\xi \xi} &=& 2 p_\ast \frac{r^2}{2} p_\ast + 2 NS^2 \left\{
\frac{(\om+1)^2}{4} + \frac{r^2}{2} \lambda (\ph^2 - v^2) \right\},\\
\lb{PB11p}
H_{a \chi} &=& 2 iNS ( (p_\ast\om) - \om p_\ast), \\
\lb{PB12p}
H_{\chi a} &=& 2 i \left\{ p_\ast NS\om + NS (p_\ast \om) \right\}, \\
\lb{PB13p}
H_{a \xi} &=& i \frac{r^2}{2} NS ((p_\ast\ph) - \ph p_\ast), \\
\lb{PB16p}
H_{\xi a} &=& i p_\ast \frac{r^2}{2} NS \ph + i \frac{r^2}{2} NS ( p_\ast \ph),
\\
\lb{PB15p}
H_{\chi\xi} &=& H_{\xi\chi} = - \ph NS^2,
\eea
where $p_\ast $ denotes the differential operator
\be
\lb{pstar}
p_\ast = - i NS \frac{d}{dr}.
\ee
Finally, let $A$ be the diagonal matrix
\be
\lb{PB18}
A = \left(
\begin{array}{lll}
N r^2 & 0 & 0 \\
0 & 2 & 0 \\
0 & 0 & r^2 \\
\end{array}
\right).
\ee
With these definitions we have the following standard form for the pulsation
equations
\be
\lb{PB19}
H \Psi = - A \ddot \Psi.
\ee
For a harmonic time dependence, $\Psi (t,r) = \Psi(r) e^{i\om t}$, this gives
the eigenvalue equation
\be
\lb{PB19a}
H \Psi = \om ^2 A \Psi.
\ee

One can show that $H$ is self-adjoint relative to the scalar product
\be
\lb{PB20}
\braket{\Psi}{\Phi} = \int_0^\infty \bar\Psi \Phi \frac{1}{NS} dr.
\ee
{}From (\ref{PB19a}) we obtain
\be
\lb{PB21}
 \om^2 = \frac{\bra{\Psi} H \, \ket{\Psi}}{\bra{\Psi} A \, \ket{\Psi}}.
\ee
For the lowest value $\om_0^2$ we have the minimum principle
\be
\lb{PB22}
\om_0^2 = \inf_\Psi  \frac{\bra{\Psi} H \, \ket{\Psi}}{\bra{\Psi} A \,
\ket{\Psi}},
\ee
where $\Psi$ runs over all functions in the domains of the operators.

We have so far left out the linearized Gauss constraint
\be
\lb{PB23}
\left( \frac{r^2}{S} \de \dot a_1\right)' =  -\frac{2}{NS} \left(\om \de \dot
\chi + \frac{r^2}{2} \ph \de \dot \xi \right).
\ee
It will turn out that it is automatically satisfied for physical pulsations
\cite{AKIB}.

\section{Instability of Regular EYMH--Solutions}
\label{Inst}
For our system, numerical solutions are given by \cite{GREE}, which can be
classified by the number of knots $k$ (zeros) of the gauge field $f$. By a
field redefinition ($\om \rightarrow -\om$), these solutions can be brought to
the form considered above: $f = \om $ and $h = \ph$.

For finite energy solutions, $\om $ and $\ph $ obey the following boundary
conditions (depending on $k$)
\bea
\lb{INST1}
&\mbox{for $k$ odd: } & \om (0) = -1 , \qquad \om (\infty ) = 1 , \nonumber\\
&& \ph (0) = 0 ,\qquad \ph (\infty ) = \pm v ; \nonumber\\ \nonumber\\
&\mbox{for $k$ even:} & \om (0) = \om (\infty ) = 1 , \nonumber\\
&& \ph(0) =\ph (\infty ) = \pm v.
\eea
The fields approach their asymptotic values exponentially.

With the following judiciously chosen one--parameter family of field
configurations
\bea
\lb{INST6}
a_{1 \alpha} &=& \alpha \om ' ,  \nonumber\\
f_\alpha &=& \frac{\om +1}{2} e^{i \alpha (\om -1) } + \frac{\om -1}{2} e^{i
\alpha (\om +1) } , \nonumber\\
h_\alpha &=& \ph e^{i \alpha \frac{\om -1}{2} },
\eea
where $\om $ and $\ph $ denote the equilibrium solutions, we shall show that
the equilibrium solutions are unstable. Obviously the family runs through the
equilibrium solution for the parameter value $\alpha = 0$.

The variations of this one--parameter family at $\alpha = 0$ are
\bea
\lb{INST7}
\de a_1 &=& \om ' , \nonumber\\
\de \chi &=& (\om ^2 - 1) ,\\
\de \xi &=& \frac{\om - 1}{2} \ph . \nonumber
\eea
These are used as trial wave functions in the minimum principle (\ref{PB21}).

The denominator $\bra{\Psi } A \ket{\Psi }$ in (\ref{PB21}) is immediately
obtained from (\ref{PB18}) and the variations (\ref{INST7}), giving
\bea
\lb{INST8a}
\nonumber\\
\bra{\Psi } A \ket{\Psi } &=& \int \left\{ \frac{r^2 (\om ')^2}{S} + 2
\frac{(\om ^2 - 1)^2}{NS} + \frac{(\om  - 1)^2 \ph ^2 r^2}{4 NS} \right\} dr.
\\
\nonumber
\eea
According to (\ref{INST1}) all terms vanish exponentially as $ r \rightarrow
\infty $, and therefore $\bra{\Psi } A \ket{\Psi } $ is finite. The special
choice of the familiy (\ref{INST6}) shows up especially  in the last term of
(\ref{INST8a}), in which the asymptotically growing factor $r^2 \ph^2$ is
damped by the coupling to the gauge field.

A direct calculation of the numerator $\bra{\Psi } H \ket{\Psi }$ for the
operator H given by (\ref{PB10p}) - (\ref{PB17p}) is somewhat tedious.
We find
\be
\lb{INST4a}
\bra{\Psi } H \ket{\Psi } = - \int \left\{ 2N(\om ')^2 + 2 \frac{(\om ^2 -
1)^2}{r^2} +  \frac{\ph ^2}{2} (\om  - 1)^2 \right\} S dr,
\ee
which is clearly negative.

This shows the existence of bound states of (\ref{PB19a}) with negative
eigenvalue $\om ^2$. Since all eigenstates with nonzero eigenvalue
automatically fulfill the Gauss constraint (\ref{PB23}), the instability of
soliton solutions is proven (see \cite{BROD2} and \cite{AKIB}).
\section{Remarks}
\label{Remarks}

This instability proof is very general in that it can be applied to a variety
of systems, including the non-abelian Proca system ("frozen Higgs field"), and,
as mentioned above, it has recently been applied to the Einstein--Yang--Mills
case for arbitrary gauge groups \cite{BROD2}.
However, it does not cover the black hole case, since $N(r_H) = 0$ for black
holes, and thus $\bra{\Psi} A \ket{\Psi}$ in (\ref{INST8a}) diverges. In fact,
it has been shown, that even the $k = 1$ Bartnick McKinnon black hole has no
imaginary direction of instability. (There remains only the "real" instability
found in ref.\ \cite{STR2}.) Furthermore, we have numerically found some of the
non--abelian Proca  black hole solutions to be  linearly stable with respect to
spherically symmetric perturbations \cite{DIPB}, \cite{DIFM}. From this, we are
tempted to conjecture the linear stability of some EYMH black holes with
respect to spherically symmetric perturbations, although we have not yet
numerically studied the problem for a dynamic Higgs field.

\end{document}